\documentclass[preprint,aps,pdftex]{revtex4-1}

\usepackage{graphicx} 
\usepackage{color}
\usepackage{tabularx}

\def\iotabar{\lower3pt\hbox{$\mathchar'26$}\mkern-8mu\iota}
    
\begin{document}
\bibliographystyle{unsrt}

\title{The impact of heating power on radial heat transport in W7-X}
\author{B.Ph.~van Milligen$^1$, U. Hoefel$^2$, M. Hirsch$^2$, B.A.~Carreras$^3$, C.~Hidalgo$^1$ and the W7-X Team}
\vspace{3em}
\affiliation{$^1$ CIEMAT - Laboratorio Nacional de Fusi{\'o}n, Avda.~Complutense 40, 28040 Madrid, Spain}
\affiliation{$^2$ Max-Planck Institute for Plasma Physics, Greifswald, Germany}
\affiliation{$^3$ Universidad Carlos III, 28911 Legan\'es, Madrid, Spain}



\maketitle

The understanding of the outward radial transport of heat in magnetic confinement fusion devices is a priority for the development of economically viable fusion reactors.

\begin{figure}\centering
  \includegraphics[trim=0 0 0 0,clip=,width=9 cm]{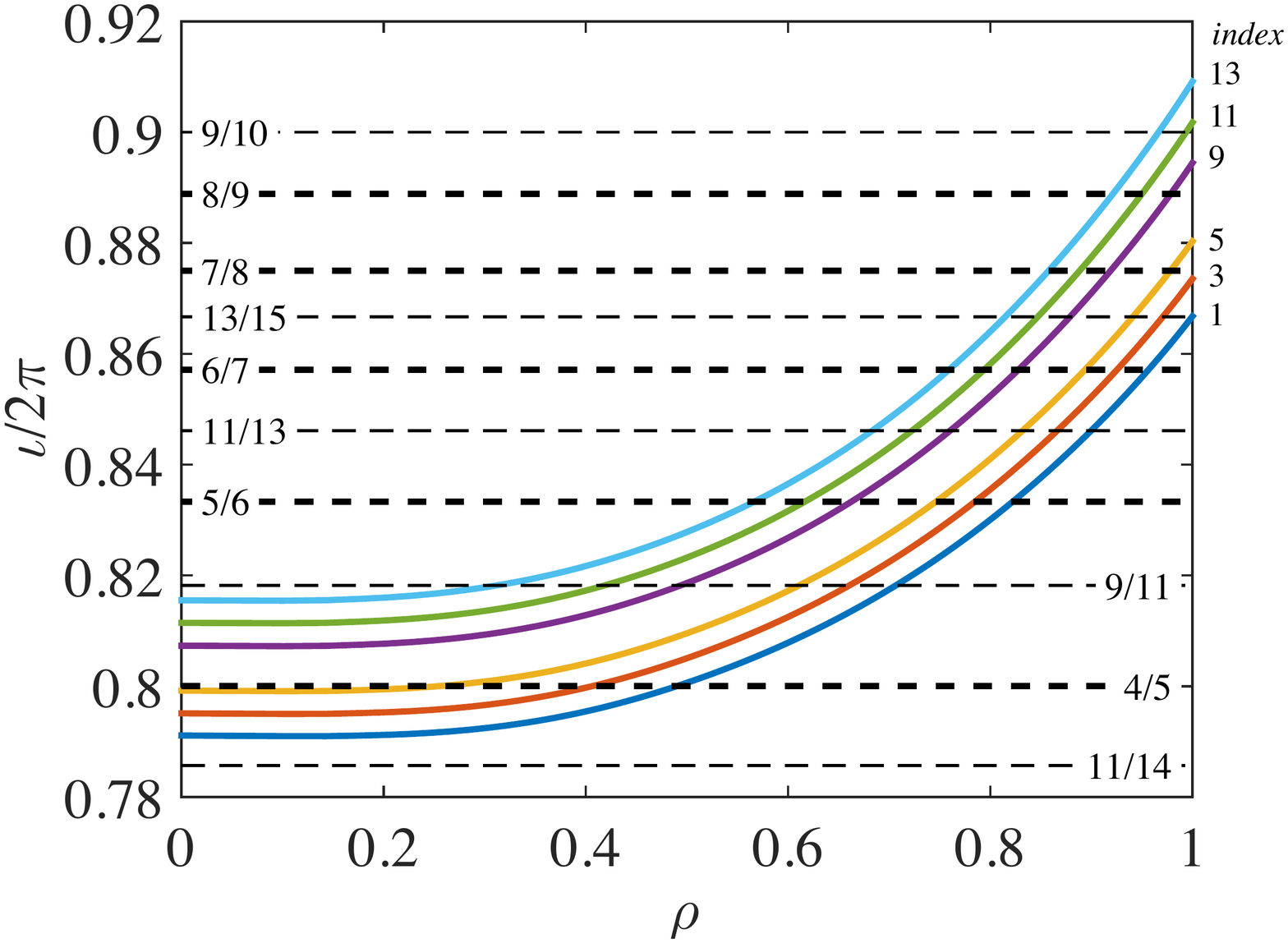}
\caption{\label{iota}Profiles of the rotational transform, $\iotabar = \iota/2\pi$ as a function of normalized radius, $\rho = r/a$, for the set of magnetic configurations studied here (identified by the index shown on the right). 
The lowest $\iotabar$ profile (index 1) corresponds to the `standard OP1.1' configuration. 
Some major rational values are indicated by horizontal dashed lines.}
\end{figure}

Here, we analyze the radial propagation of spontaneously generated electron temperature ($T_e$) fluctuations~\cite{Sattler:1994} measured using the Electron Cyclotron Emission (ECE) diagnostic in Wendelstein 7-X, which disposes of 32 channels, covering a large part of the plasma minor radius~\cite{Marushenko:2006,Koenig:2015}.
Wendelstein 7-X is a helical advanced stellarator (HELIAS) with major radius $R = 5.5$ m, minor radius $a \simeq 0.5$ m, 5 field periods, and a toroidal magnetic field of $B \simeq 3$ T.
The design of the device was based, among others, on the optimization of Neoclassical particle transport and MHD stability~\cite{Wanner:2001,Wolf:2008}.
The discharges analyzed were characterized by a low line average electron density (${\overline n_e} \simeq 1.5 \cdot 10^{19}$ m$^{-3}$)
and were heated by Electron Cyclotron Resonance Heating (ECRH)~\cite{Sattler:1994,White:2008}, with up to about 4 MW of power~\cite{Klinger:2017}, resulting in so-called Core Electron Root Confinement (CERC) plasmas~\cite{Dinklage:2016}.
We analyzed a set of discharges with different rotational transform profiles ($\iotabar=\iota/2\pi$), shown in Fig.~\ref{iota}. 

The analysis is based on a relatively new technique known as the Transfer Entropy (TE)~\cite{Schreiber:2000}.
In the present context, the main relevant feature of this technique is that it allows exploiting the propagation of small, randomly occurring temperature fluctuations to effectively probe heat transport~\cite{Milligen:2017}.
This converts the TE into a valuable technique to explore heat transport directly, as it is non-perturbative, contrasting with the commonly used power modulation technique that involves a significant external perturbation~\cite{Jacchia:1991}.
The TE is a measure of the causal relation or information flow between two signals $Y$ and $X$: it quantifies the number of bits by which the prediction of a signal $X$ can be improved by using the time history of not only the signal $X$ itself, but also that of signal $Y$.

We use a simplified version of the Transfer Entropy, calculated as follows from discretely sampled time series data $x_i$ and $y_j$, corresponding to signals $X$ and $Y$, respectively:
\begin{eqnarray}\label{TE}
T_{Y \to X} =& \sum{p(x_{n+1},x_{n-k},y_{n-k}) \times }\nonumber \\ 
& \log_2 \frac{p(x_{n+1}|x_{n-k},y_{n-k})}{p(x_{n+1}|x_{n-k})} .
\end{eqnarray}
Here, $p(a|b)$ is the probability distribution of $a$ conditional on $b$, $p(a|b) = p(a,b)/p(b)$.
The probability distributions $p(a,b,c,\dots)$ are constructed using $m$ bins for each argument, i.e., the object $p(a,b,c,\dots)$ has $m^d$ bins, where $d$ is the dimension (number of arguments) of $p$. 
The sum in Eq.~\ref{TE} runs over the corresponding discrete bins.
The number $k$ can be converted to a `time lag' by multiplying it by the sampling rate.
The construction of the probability distributions is done using `course graining', i.e., a low number of bins (here, $m=3$), to obtain statistically significant results. For more information on the technique, please refer to Ref.~\cite{Milligen:2014}. 

Of course, `propagation of information' is not the same as the `propagation of heat pulses', as the former does not depend on signal amplitude, while the latter does.
On the other hand, (heat) diffusion is mainly a geometric property -- the spreading in time of an initially localized perturbation -- that can be explored effectively using the Transfer Entropy, calculated from temperature measurements.

To visualize radial propagation (of information encoded in the $T_e$ fluctuations), we have calculated the Transfer Entropy in a number of discharges with different conditions, between a reference ECE channel and all other available ECE channels. The ECE sampling rate was 2 MHz.
Within each discharge, the ECRH heating power was varied.
An initial `high power' phase (ECRH power $P_{\rm ECRH} \simeq 2.0$ MW) was followed by a `low power' phase ($P_{\rm ECRH} \simeq 0.6$ MW) and a `medium power' phase ($P_{\rm ECRH} \simeq 1.3$ MW), each phase lasting about 0.3 s.
The discharges correspond to different magnetic configurations and $\iotabar$ profiles (cf.~Fig.~\ref{iota}), such that rational surfaces are placed at different radial  locations (as indicated in the figures).
The vertical axes and rational surfaces shown in the figures always correspond to the nominal $\rho$ values of the ECE channels, corrected for  the Shafranov shift. 
Negative values of $\rho$ correspond to the low field side of the plasma, and positive values to the high field side.

\begin{figure}\centering
  \includegraphics[trim=50 0 400 0,clip=,width=9cm]{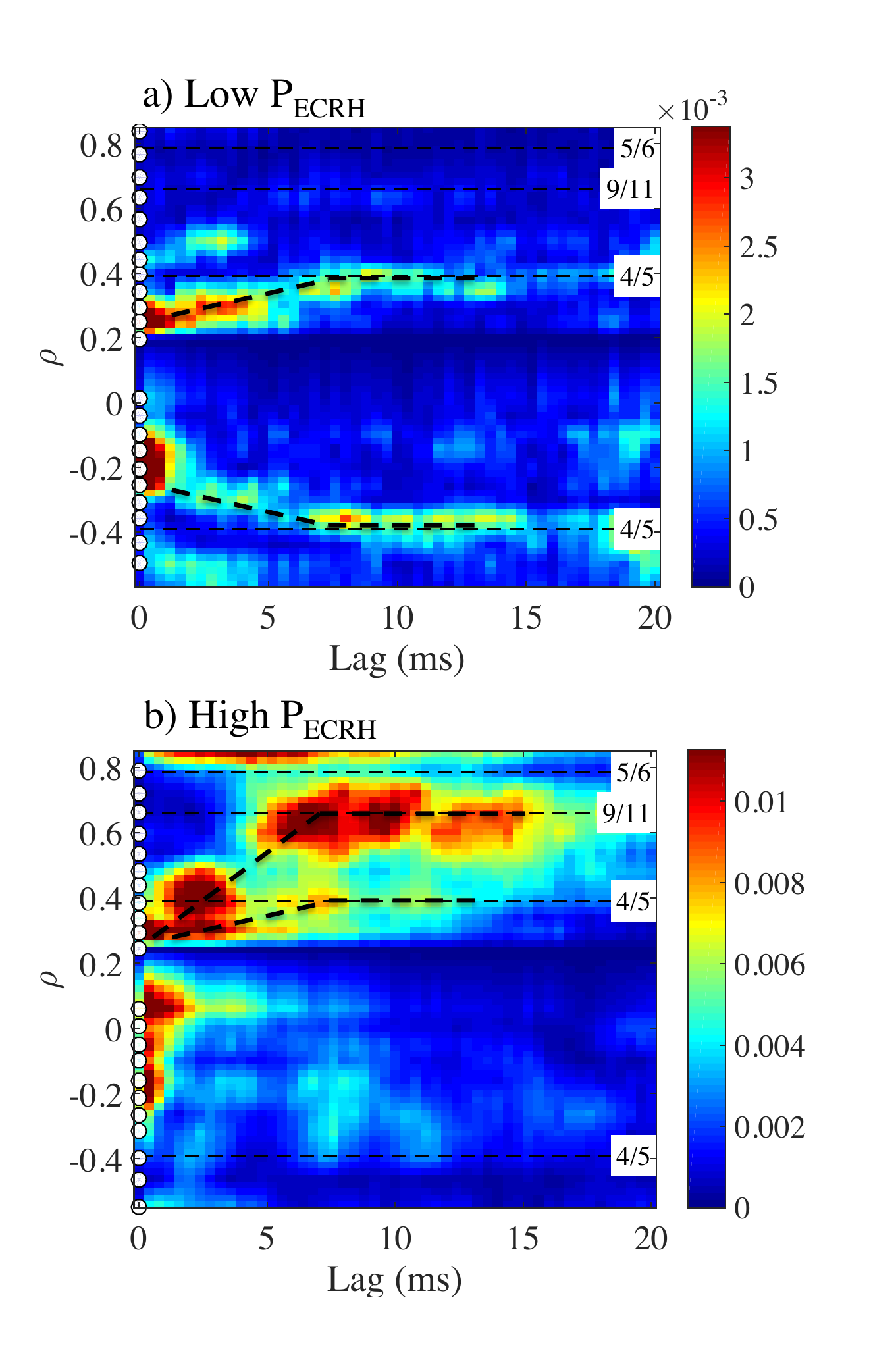}
\caption{\label{20160309_016_hi_lo}Transfer Entropy vs.~$\rho$ and time lag, $\tau$, for discharge 016, date 2016-03-09. Configuration index: 3.
Dots indicate the location of ECE measurement channels.
Horizontal dashed lines indicate the location of rational surfaces. 
Radial propagation is indicated with thick dashed lines.}
\end{figure}

We are most interested in outward propagation. Therefore, we use a reference channel located close to $\rho = 0.2$, not far outside the central ECRH power deposition region. 
In each power phase, we use time intervals with lengths between 0.22 and 0.3 s for the analysis.
Fig.~\ref{20160309_016_hi_lo} shows an example of the Transfer Entropy versus time lag, $\tau$, and normalized radius, $\rho$.
Comparing the low and high ECRH power phases, one observes that they have in common that some perturbations propagate outward relatively slowly to the 4/5 rational surface, which acts as a `trapping zone' for these perturbations.
In the high power phase, there is an additional branch of radial propagation, faster and more intense (in terms of information transfer), reaching the 9/11 rational surface.
This suggests that the larger perturbations in the high power phase may either propagate faster or achieve mode coupling with perturbations near the 9/11 rational surface.  

The figure shows that propagation is not continuous, but experiences delays (`trapping') at certain radial positions, which appear to be associated with low-order rational surfaces, within the available resolution.
Also, apparent propagation `jumps' occur, when the response at a given outward location occurs at a smaller time lag than at some other locations further inward.
In particular, two transport branches may be discerned in Fig.~\ref{20160309_016_hi_lo}:
(a) relatively slow outward propagation up to the 4/5 rational surface, visible in both the low and high ECRH power phases, and
(b) much faster radial propagation, apparently discontinuous, reaching the 9/11 rational surface, only visible in the high power phase.
This result seems to suggest that an additional transport channel is activated at increased heating power, reminiscent of `critical gradient' transport models~\cite{Ryter:2006}.
This effect is likely related to the phenomenon of `power degradation' (see below).

To quantify the time evolution of the centre of gravity of the propagating perturbations, we calculate the mean radius of the propagating information as follows:
\begin{equation}
\langle \rho \rangle = \frac{\int{T_{\rho_0\to \rho} \rho d\rho } }{ \int{T_{\rho_0\to \rho} d\rho } },
\end{equation}
i.e., the weighted mean of the radius, using the Transfer Entropy $T_{\rho_0\to \rho}$ between the reference position and position $\rho$ as the weight.
The integration only includes positive values of $\rho$.
This quantity is evaluated for each value of the lag $\tau$.

\begin{figure}\centering
  \includegraphics[trim=0 0 0 0,clip=,width=9cm]{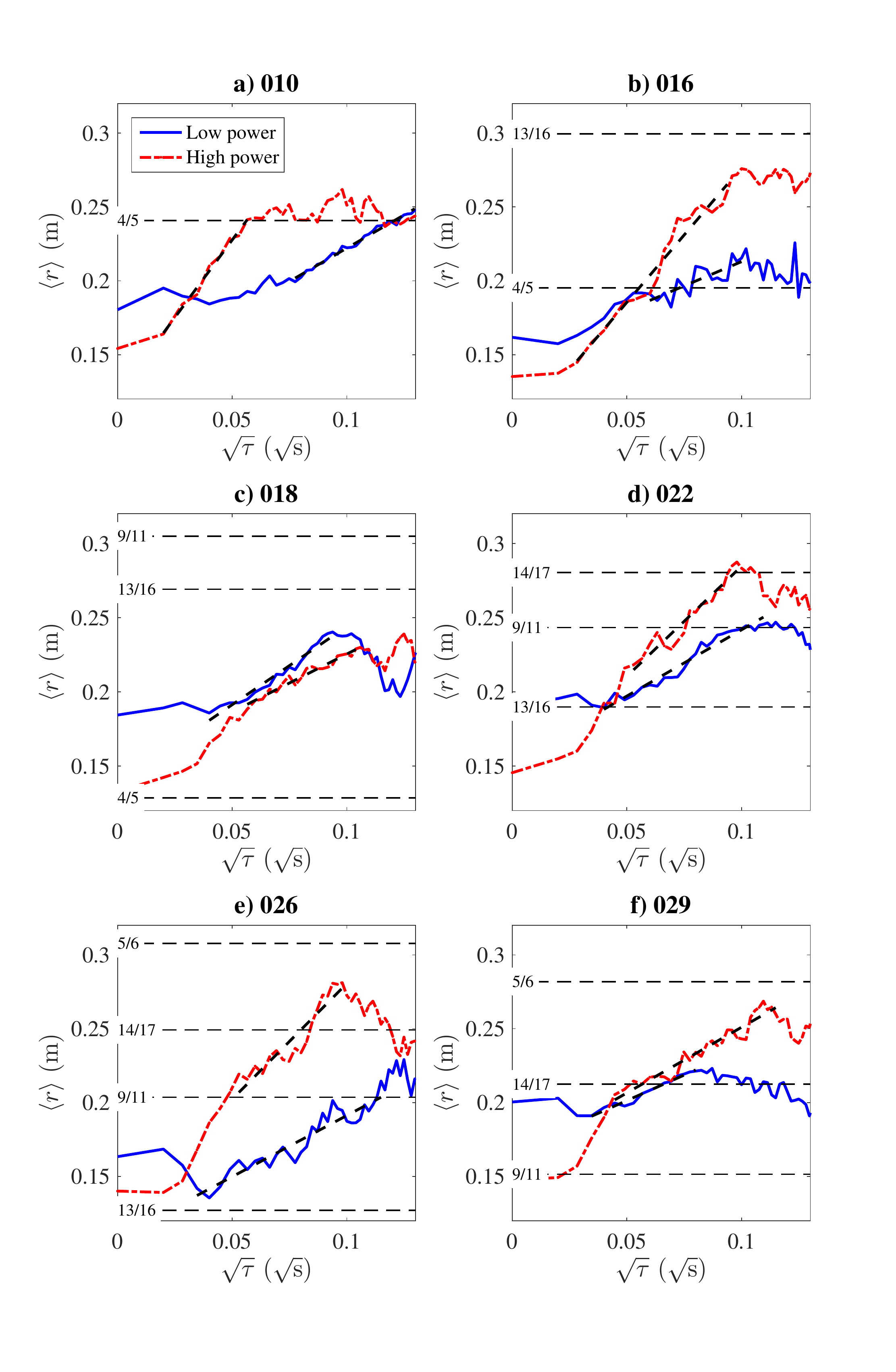}
\caption{\label{chi_eff}$\langle r \rangle = \langle \rho \rangle a$ as a function of $\sqrt{\tau}$ for various discharges with different $\iotabar$ at two different heating levels, $P_{\rm ECRH}$ (date of experiment: 2016-03-09).
Horizontal dashed lines indicate the location of some low-order rational surfaces.
Slanted dashed lines indicate linear fits made to estimate the local effective heat diffusivity (see Table~\ref{table_chi}).
}
\end{figure}

Fig.~\ref{chi_eff} shows results for $\langle r \rangle = \langle \rho \rangle a$ as a function of $\sqrt{\tau}$. 
Values at small lags ($\sqrt{\tau}<0.03$) should be ignored, as these are dominated by power deposition effects, rather than propagation.
The slope of the curves is a measure of $\sqrt{\chi_{\rm eff}}$, where $\chi_{\rm eff}$ is the effective heat diffusivity.
To obtain specific estimates of $\chi_{\rm eff}$, the curves were fitted in a range of values of $\sqrt{\tau}$, resulting in the dashed fit lines.
The length of these fit lines reflects the range of values used and typically corresponds to the interval of $\tau$ (and $\rho$) values for which significant propagation is visible in the TE graphs; for example, in discharge 016, the fits were made up to about 9 ms (cf.~Fig.~\ref{20160309_016_hi_lo}).
This also means that the obtained $\chi_{\rm eff}$ values are {\it local} rather than global and correspond roughly to $r \simeq a/2$.
Table~\ref{table_chi} lists the values of $\chi_{\rm eff}$ obtained from the linear fits shown in the figure (dashed lines).

\begin{table}[htp]
\caption{Effective local heat diffusivity $\chi_{\rm eff}$ (at $r \simeq a/2$) for low and high ECRH power, from the fits shown in Fig.~\ref{chi_eff}.}
\begin{center}
\begin{tabular}{|c|c|c|c|}
\hline
Discharge & $\iotabar$-index & $\chi_{\rm eff}^{\rm lo}$ (m$^2$/s) & $\chi_{\rm eff}^{\rm hi}$ (m$^2$/s) \\
\hline
010 & 1 & $0.81 \pm 0.08$  & $4.41 \pm 0.98$ \\
016 & 3 & $0.43 \pm 0.37$ & $3.33 \pm 0.68$ \\
018 & 5 & $1.11 \pm 0.20$ & $0.61 \pm 0.13$ \\
022 & 9 & $0.80 \pm 0.14$ & $2.20 \pm 0.65$ \\
026 & 11 & $0.69 \pm 0.18$ & $2.46 \pm 0.99$ \\
029 & 13 & $0.48 \pm 0.10$ & $0.79 \pm 0.17$ \\
\hline
\end{tabular}
\end{center}
\label{table_chi}
\end{table}%

The results clarify several things:
(1) $\chi_{\rm eff}$ is not constant in radius, but varies considerably and has a tendency to drop near major rational surfaces (indicated by horizontal dashed lines) -- the `trapping zones' noted above.
(2) Generally speaking, $\chi_{\rm eff}$ is considerably higher with high ECRH power than with low ECRH power. As is evident from the TE shown in Fig.~\ref{20160309_016_hi_lo}, this is at least partly due to information `jumping' radially (coupling effects).
(3) The lowest values of the deduced local $\chi_{\rm eff}$ are consistent with the mean global heat transport coefficient ($\chi_{\rm global} \simeq 0.3$ m$^2$/s) deduced from ECRH modulation experiments~\cite{Hoefel:2016}. 

To summarize: in this work, we use the Transfer Entropy to study the radial transport of heat.
We find that 
(a) transport is not smooth and continuous (as would be the case for purely diffusive/collisional transport), but involves alternate `jumps' (phases of rapid transport) and `trappings' (phases of slow transport); at least part of the heat transport occurs therefore in a stepwise rather than continuous fashion,
(b) these `trapping zones' or minor transport barriers, where radial transport is reduced, appear to be associated with rational surfaces,
(c) the rapid transport phases have effective transport coefficients $\chi_{\rm eff}$ of the order of or significantly larger (by an order of magnitude) than the global heat diffusivity, and   
(d) increased heating power leads to increased effective heat diffusivity due to an increase of the jump size and/or speed, 
cf.~ Fig.~\ref{20160309_016_hi_lo}, depending on the magnetic configuration.
The mentioned `jumps' possibly involve mode coupling effects and may be related to a phenomenon called `non-local' transport in literature~\cite{Gentle:1995,Kissick:1998,Mantica:1999}). 

Estimates of the heat diffusivity based on the Transfer Entropy results indicate that at high power, the outward propagation between the minor transport barriers is typically faster than expected from the global heat diffusivity ($\chi_e \simeq 0.3$ m$^2$/s~\cite{Hoefel:2016}). 
Hence, in these ECRH discharges, the minor transport barriers play a vital role in achieving the good global transport properties of \mbox{W7-X}. 

The discussed power scan experiments show that a faster transport channel may be activated when the power is increased, which provides a completely new insight into the ubiquitous phenomenon of power degradation in fusion plasmas, by which radial heat transport increases when the externally applied heating power $P$ is raised, such that the energy confinement time scales as $\tau_E \propto P^{-0.6}$, approximately~\cite{Stroth:1996b,ITER:1999,Carreras:1997,Hirsch:2008,Dinklage:2007}.
The activation of faster transport at higher heating power is a reflection of plasma self-organization.
Traditionally, this phenomenon is explained by assuming that the electron heat diffusivity $\chi_e$ would depend in a non-linear fashion on, e.g., the electron temperature gradient, $\nabla T_e$, i.e., a strictly local dependence of transport~\cite{Lopes:1995}, which can be handled in the framework of diffusive modeling (although previous work already suggested that local modeling cannot explain all phenomena~\cite{Stroth:1996b,Ida:2015}).
The present study clearly suggests that power degradation may be due to the enhanced `jump' size or speed and/or the breakdown of some of the minor transport barriers.
This implies a radical departure from the localist view, as it suggests an important role for long-range effects.

More generally, this work confirms earlier studies that showed convincingly that rational surfaces have a significant impact on radial heat transport and may in fact be essential in setting the global energy confinement, e.g., in RTP~\cite{Lopes:1997} and Alcator C-Mod~\cite{Wukitch:2002}.
Recently, a study similar to the present one was performed at TJ-II, with similar conclusions although less resolution~\cite{Milligen:2017}.

By applying a relatively new statistical analysis technique (the Transfer Entropy), the present work provides, for the first time, a view of the detailed mechanism of electron heat transport in fusion plasmas.
The reported observations provide support for various explanatory concepts suggested in literature: namely, critical gradients, non-locality, and self-organization.
Modeling the observed behavior likely requires looking beyond the (local) diffusion paradigm.
Future work may involve extending this analysis to other magnetic configurations and/or heating scenarios, including, e.g., Neutral Beam Heating.

\section*{Acknowledgements}
Research sponsored in part by the Ministerio de Econom\'ia y Competitividad of Spain under project 
Nr. 
ENE2015-68206-P. 
This work has been carried out within the framework of the EUROfusion Consortium and has received funding from the Euratom research and training programme 2014-2018 under grant agreement No 633053. 
The views and opinions expressed herein do not necessarily reflect those of the European Commission.



\end{document}